\documentclass[preprint,endfloats,prb,onecolumn,final,showpacs]{revtex4}%
\usepackage{graphicx,amsmath,amsthm,amssymb}
\usepackage{amsfonts}
\usepackage{graphicx}
\usepackage{epsfig,subfigure}
\usepackage{amsmath}
\usepackage{amssymb}%
\makeatletter
\def\@dotsep{4.5}
\makeatother
\begin{document}
\title{A Coherent Nonlinear Optical Signal Induced by Electron Correlations}
\author{Shaul Mukamel$^{1}$, Rafa{\l } Oszwa{\l }dowski$^{2,1}$, and Lijun Yang$^{1} $}
\affiliation{$^{1}$ Department of Chemistry, University of California, Irvine, CA 92697,}
\affiliation{$^{2}$ Instytut Fizyki, Uniwersytet Miko{\l }aja Kopernika, Grudzi\c{a}dzka
5/7, 87-100, Toru\'n, Poland }

\begin{abstract}
The correlated behavior of electrons determines the structure and optical
properties of mo\-le\-cu\-les, semiconductor and other systems. Valuable
information on these correlations is provided by measuring the response to
femtosecond laser pulses, which probe the very short time period during which
the excited particles remain correlated. The interpretation of
four-wave-mixing techniques, commonly used to study the energy levels and
dynamics of many-electron systems, is complicated by many competing effects
and overlapping resonances. Here we propose a coherent optical technique,
specifically designed to provide a background-free probe for electronic
correlations in many-electron systems. The proposed signal pulse is generated
only when the electrons are correlated, which gives rise to an extraordinary
sensitivity. The peak pattern in two-dimensional plots, obtained by displaying
the signal vs. two frequencies conjugated to two pulse delays, provides a
direct visualization and specific signatures of the many-electron wavefunctions.

\end{abstract}

\pacs{71.35.Cc, 73.21.La, 78.47.+p}
\maketitle

Predicting the energies and wavefunctions of interacting electrons lies at the
heart of our understanding of all structural, optical, and transport
properties of molecules and materials
\cite{Giuliani,Bartlett,Roos,Marques,Fulde,Wilson,Lee,Kotliar,Sherrill,Curtiss}%
. The Hartree-Fock (HF) approximation provides the simplest description of
interacting fermions \cite{Giuliani,Fulde}. At this level of theory each
electron moves in the average field created by the others. This provides a
numerically tractable, uncorrelated-particle picture for the electrons, that
approximates many systems well and provides a convenient basis for higher-
level descriptions. Electronic dynamics is described in terms of orbitals, one
electron at a time. Correlated n-electron wavefunctions, in contrast, live in
a high (3n) dimensional space and may not be readily visualized. Deviations
from the uncorrelated picture (correlations), are responsible for many
important effects. Correlation energies are comparable in magnitude to
chemical bonding energies and are thus crucial for predicting molecular
geometries and reaction barriers and rates with chemical accuracy. These
energies can be computed for molecules by employing a broad arsenal of
computational techniques such as perturbative corrections \cite{Curtiss},
configuration interaction \cite{Sherrill}, multideterminant techniques
\cite{Roos}, coupled cluster theory \cite{Bartlett} and time dependent density
functional theory TDDFT \cite{Giuliani,Marques,Onida}. Correlation effects are
essential in superconductors \cite{Lee,Kotliar,Elsaesser} and can be
manipulated in artificial semiconductor nanostructures
\cite{Klimov,Rossi,ChemlaShah2001}. The fields of quantum computing and
information are based on manipulating correlations between spatially separated
systems, this is known as entanglement \cite{Qcomputing}. \ 

In this article we propose a nonlinear optical signal that provides a unique
probe for electron correlations. The technique uses a sequence of three
optical pulses with wavevectors $\mathbf{k}_{1}\mathbf{,k}_{2}$ and
$\mathbf{k}_{3}$, and detects the four wave mixing signal generated in the
direction $\mathbf{k}_{S}\mathbf{=k}_{1}\mathbf{+k}_{2}\mathbf{-k}_{3}$ by
mixing it with a fourth pulse (heterodyne detection).\cite{mukbook} \ We show
that this correlation-induced signal $S_{CI}(t_{3},t_{2},t_{1})$, which
depends parametrically on the consequentive delays $t_{1},$ $t_{2},$ $t_{3}$
between pulses, vanishes for uncorrelated systems, providing a unique
indicator of electron correlations. This technique opens up new avenues for
probing correlation effects by coherent ultrafast spectroscopy.

Starting with the HF ground-state ($g$) of the system, each interaction with
the laser fields can only move a single electron from an occupied to an
unoccupied orbital. The first interaction generates a manifold ($e$) of single
electron-hole (e-h) pair states. A second interaction can either bring the
system back to the ground state or create a second e-h pair. We shall denote
the manifold of doubly excited states as $f$ (Fig.~1). We can go on to
generate manifolds of higher levels. However, this will not be necessary for
the present technique. The quantum pathways (i) and (ii) contributing to this
signal can be represented by the Feynman diagrams \cite{mukbook} shown in
Fig.~1. Each diagram shows the sequence of interactions of the system with the
various fields and the state of the electron density matrix during each delay
period. We shall display the signal as $S_{CI}(\Omega_{3},\Omega_{2},t_{1})$,
where $\Omega_{3},$ $\Omega_{2}$ are frequency variables conjugate to the
delays $t_{3}$ and $t_{2}$ (Fig.~1) by a Fourier transform%
\[
S_{CI}(\Omega_{3},\Omega_{2},t_{1})=\int_{0}^{\infty}\int_{0}^{\infty}%
dt_{2}dt_{3}S_{CI}(t_{3},t_{2},t_{1})\exp(i\Omega_{2}t_{2}+i\Omega_{3}t_{3}),
\]
with $t_{1}$ fixed. This yields an expression for the exact response function%
\begin{align}
&  S_{CI}(\Omega_{3},\Omega_{2},t_{1}=0)\label{Eq.SOS_resp}\\
&  =\sum_{e,e^{\prime},f}\frac{1}{\Omega_{2}-\omega_{fg}}\left[  \frac
{\mu_{ge}\mu_{ef}\mu_{fe^{\prime}}\mu_{e^{\prime}g}}{\Omega_{3}-\omega
_{e^{\prime}g}}-\frac{\mu_{ge^{\prime}}\mu_{e^{\prime}f}\mu_{fe}\mu_{eg}%
}{\Omega_{3}-\omega_{fe}}\right]  ,\nonumber
\end{align}
where for simplicity we set $t_{1}=0$. Two-dimensional correlation plots of
$\Omega_{2}$ vs. $\Omega_{3}$ then reveal a characteristic peak pattern, which
spans the spectral region permitted by the pulse bandwidths. The two terms in
the brackets correspond respectively to diagrams (i) and (ii) of Fig.~1. Here
$\mu_{\nu\nu^{\prime}}$ are the transition dipoles and $\omega_{\nu\nu
^{\prime}}$ are the transition energies between electronic states, shifted by
the pulse carrier frequency $\omega_{0}$, i.e., $\omega_{e^{\prime}g}%
=\epsilon_{e^{\prime}}-\epsilon_{g}-\omega_{0},$ $\omega_{fe}=\epsilon
_{f}-\epsilon_{e}-\omega_{0}$ and $\omega_{fg}=\epsilon_{f}-\epsilon
_{g}-2\omega_{0}$. This shift eliminates the high optical frequencies. The
carrier frequency of the three beams, $\omega_{0}$, is held fixed and used to
select the desired spectral region. In Eq.~(\ref{Eq.SOS_resp}) we have invoked
the rotating wave approximation (RWA), and only retained the dominant terms
where all fields are resonant with an electronic transition.

In both diagrams, during $t_{2}$ the system is in a coherent superposition
(coherence) between the doubly excited state $f$ and the ground state $g$.
This gives the common prefactor $(\Omega_{2}-\omega_{fg})^{-1}$. As
$\Omega_{2}$ is scanned, the signal will thus show resonances corresponding to
the different doubly-excited states $f$. However, the projection along the
other axis $(\Omega_{3})$ is different for the two diagrams. In diagram (i)
the system is in a coherence between $e^{\prime}$ and $g$ during $t_{3}$. As
$\Omega_{3}$ is scanned, the first term in brackets reveals single excitation
resonances when $\Omega_{3}=\omega_{e^{\prime}g}$. \ For the second diagram
(ii) the system is in a coherence between $f$ and $e^{\prime}$ during $t_{3}$.
This gives resonances at $\Omega_{3}=\omega_{fe}$ in the second term in the
brackets. \ Many new peaks corresponding to all possible transitions between
doubly and singly excited states $\omega_{fe}$ should then show up. \ 

The remarkable point that makes this technique so powerful is that the two
terms in Eq.~(\ref{Eq.SOS_resp}) interfere in a very special way. For
independent electrons, where correlations are totally absent, the two e-h pair
state $f$ is simply given by a direct product of the single pair states $e$
and $e^{\prime}$, and the double-excitation energy is the sum of the
single-excitation energies $\epsilon_{f}=\epsilon_{e}+\epsilon_{e^{\prime}}$,
so that $\omega_{e^{\prime}g}=\omega_{fe}$ and the two terms in the brackets
exactly cancel. Density functional theory \cite{Giuliani,Marques,Onida}, when
implemented using the Kohn-Sham approach, gives a set of orbitals that carry
some information about correlations in the ground state. \ The signal
calculated using transitions between Kohn-Sham orbitals will vanish as well.
\ This will be the case for any uncorrelated-particle calculation that uses
transitions between fixed orbitals, no matter how sophisticated was the
procedure used to compute these orbitals. \ We expect the resonance pattern of
the two-dimensional $S_{CI}$ signal to provide a characteristic fingerprint
for electron correlations. \ 

The following simulations carried out for simple model systems which contain a
few orbitals and electrons, illustrate the power of the proposed technique.
\ Doubly excited states can be expressed as superpositions of products of two
e-h pair states. \ Along $\Omega_{3}$ we should see the various doubly excited
states at $\omega_{fg}$, whereas along $\Omega_{2}$ we observe the various
projections of the $f$ state onto single pair states $\omega_{e^{\prime}g}$
and the differences $\omega_{fe}=\omega_{fg}-\omega_{eg}$. \ The 2D spectra
thus provide direct information about the nature of the many body
wavefunctions that is very difficult to measure by any other means. The
patterns predicted by different levels of electronic structure simulations
provide a direct means for comparing their accuracy. \ We used a tight-binding
Hamiltonian $H=H_{0}+H_{C}+H_{L}$. The single-particle contribution $H_{0}$
contains orbital energies and hoppings%
\[
H_{0}=\sum_{m_{1},n_{1}}t_{m_{1},n_{1}}c_{m_{1}}^{\dagger}c_{n_{1}}%
+\sum_{m_{2},n_{2}}t_{m_{2},n_{2}}d_{m_{2}}^{\dagger}d_{n_{2}},
\]
where $c_{n_{1}}$ and $d_{n_{2}}$ are electron and hole annihilation operators
respectively, and the summations run over spin-orbitals. We assume equal
hopping $t$ for electrons and holes. The many-body term responsible for
correlations%
\[
H_{C}=\frac{1}{2}\sum_{m_{1},n_{1}}V_{m_{1}n_{1}}^{ee}c_{m_{1}}^{\dagger
}c_{n_{1}}^{\dagger}c_{n_{1}}c_{m_{1}}+\frac{1}{2}\sum_{m_{2},n_{2}}%
V_{m_{2}n_{2}}^{hh}d_{m_{2}}^{\dagger}d_{n_{2}}^{\dagger}d_{n_{2}}d_{m_{2}%
}-\sum_{m_{1},m_{2}}V_{m_{1}m_{2}}^{eh}c_{m_{1}}^{\dagger}d_{m_{2}}^{\dagger
}d_{m_{2}}c_{m_{1}},
\]
contains only direct Coulomb couplings. The electron-electron, hole-hole and
electron-hole interaction are denoted $V^{ee},$ $V^{hh},$ and $V^{eh}$,
respectively. \ Values of $t$ and Coulomb integrals $V_{00}^{eh}$,
$V_{01}^{eh}$ (subscripts $0$ and $1$ indicate the sites) were derived by
fitting emission spectra of coupled quantum dots\cite{Baye2001}. Owing to the
nature of quantum dots states, we can assume that $V^{eh}\simeq V^{ee}\simeq
V^{hh}$.\cite{Baye2000} We choose $V^{ee}=V^{hh}=1.2V^{eh}$ and use these
values for all orbitals. $\ H_{L}$ describes the dipole interaction with the
laser pulses, $H_{L}=-E(t)\mu_{m_{1}m_{2}}d_{m_{2}}c_{m_{1}}+h.c$, where
$E(t)$ is the light field and $\mu_{m_{1}m_{2}}$ are local dipole moments of
various orbitals $m_{1}$, $m_{2}$. \ $\omega_{0}$ was tuned to the
single-particle optical gap energy (1~3 eV). \ 

Even though our parameters are fitted to QDs, the overall picture emerging
from the calculations can be applied to a wider class of systems, whose
optical response is determined by correlated e-h pairs. We have employed an
equation of motion approach for computing the signal. Many-body states are
never calculated explicitly in this algorithm. Instead, we obtain the signal
directly by solving the Nonlinear Exciton Equations
(NEE)\cite{ChernyakZhangMukamel98,AxtMukamel1998rev}. These equations describe
the coupled dynamics of two types of variables representing single e-h pairs:
$B_{m}=\left\langle d_{m_{2}}c_{m_{1}}\right\rangle $ (here $m=\left(
m_{1},m_{2}\right)  $ stands for both the electron index $m_{1}$ and the hole
index $m_{2}$) and two pairs $Y_{mn}=\left\langle d_{m_{2}}c_{m_{1}}d_{n_{2}%
}c_{n_{1}}\right\rangle $. \ For our model the NEE is equivalent to full CI,
and yield the exact signal with all correlation effects fully included. \ This
signal provides a direct experimental test for many-body theories, which use
various degrees of approximations to treat electron correlations. We compare
the exact calculation (NEE) with the time dependent Hartree Fock (TDHF)
theory, which is an approximate, widely used technique for treating
correlations by factorizing the $Y$ variables into $\left\langle d_{m_{2}%
}c_{m_{1}}\right\rangle \left\langle d_{n_{2}}c_{n_{1}}\right\rangle
-\left\langle d_{n_{2}}c_{m_{1}}\right\rangle \left\langle d_{m_{2}}c_{n_{1}%
}\right\rangle $. This assumes that two e-h pairs are independent and we only
need to solve the equations for $\left\langle d_{n_{2}}c_{n_{1}}\right\rangle
$. Correlation within e-h pairs is nevertheless retained by this level of
theory, as evidenced by the finite $S_{CI}$ signal. The equations of motion
derived using both levels of factorization are solved analytically, yielding
the exact and the TDHF $S_{CI}$ signals. The TDHF solutions have the following
structure: a set of single-particle excitations with energies $\epsilon
_{\alpha}$ and the corresponding transition dipole moments are obtained by
solving the linearized TDHF equations. Many-particle state energies are given
by sums of these elementary energies. \ Two-particle energies are of the form
$\epsilon_{\alpha}+\epsilon_{\beta}$. This approximation is the price we pay
for the enormous simplicity and convenience of TDHF. Correlated many-electron
energies computed by higher level techniques do not possess this additivity property.

In general the TDHF signal contains a different number of resonances along
$\Omega_{2}$ than the exact one. Their positions, $\omega_{fg}=\epsilon
_{f}-\epsilon_{g},$ also differ since the former uses the additive
approximation for the energies $\epsilon_{f}.$ The $\Omega_{3}$ value of each
resonance in the exact simulation is given by either $\omega_{e^{\prime}g}$
(first term in brackets in Eq.~\ref{Eq.SOS_resp}) or $\omega_{fe}$ (second
term in the brackets). The simulations of the TDHF response function presented
below show fewer peaks than in the exact calculation. This dramatic effect
reflects direct signatures of the correlated two e-h pair wavefunction, which
are only revealed by the $S_{CI}$ technique.

We first consider a simple model, consisting of a single site with one valence
orbital and one conduction orbital (Fig.~2A). The energy of the
(spin-degenerate) single-pair state is $\epsilon_{e}=-V^{eh}$. The only
two-pair state has energy $\epsilon_{f}=-4V^{eh}+V^{ee}+V^{hh}$, compared with
$\bar{\epsilon}_{f}=2\epsilon_{e}$ in the TDHF approximation (the TDHF
double-excited energies and frequencies will be marked with a bar:
$\bar{\epsilon}_{f}$, $\bar{\omega}_{fe}$). Thus, the exact signal has two
peaks at $(\Omega_{3},\Omega_{2})=\left(  \omega_{eg},\omega_{fg}\right)  $
and $\left(  \omega_{fe},\omega_{fg}\right)  $, while TDHF predicts only one
peak at $(\Omega_{3},\Omega_{2})=\left(  \omega_{eg},\bar{\omega}_{fg}\right)
=\left(  \bar{\omega}_{fe},\bar{\omega}_{fg}\right)  $.

This high sensitivity to correlation effects is general and is maintained in
more complex systems. In Fig.~2B we consider a system with two valence
orbitals with a splitting $\Delta$ and one conduction orbital. It has 2 single
e-h pair transitions $e_{1},$ $e_{2}$ with energies $\epsilon_{1}=-V^{eh}$ and
$\epsilon_{2}=-V^{eh}+\Delta$. The exact spectrum contains 8 peaks, with
$\epsilon_{f}$ energies being sums of all quasiparticle interactions and hole
level energies: $\epsilon_{f_{1}}=-4V^{eh}+V^{ee}+V^{hh}$, $\epsilon_{f_{2}%
}=\epsilon_{f_{1}}+\Delta$, $\epsilon_{f_{3}}=\epsilon_{f_{1}}+2\Delta$.
Within TDHF we find $\bar{\epsilon}_{f_{1}}=2\epsilon_{1},$ $\bar{\epsilon
}_{f_{2}}=\epsilon_{1}+\epsilon_{2}$,$\ \bar{\epsilon}_{f_{3}}=2\epsilon_{2}$
and the 2D spectrum only shows 4 peaks.

The simple energy-level structure of the two systems (Fig.~2A,B), whereby
$\bar{\omega}_{fe}=\omega_{eg}$, allows an insight into the differences in
predictions of the two response functions. We recast Eq.~(\ref{Eq.SOS_resp})
for the exact signal in a slightly different form: $\left(  \Omega_{2}%
-\omega_{fg}\right)  ^{-1} \left(  \Omega_{3}-\omega_{e^{\prime}g}\right)
^{-1} \left(  \Omega_{3}-\omega_{fe}\right)  ^{-1} $ (all dipoles for this
system are equal $\mu_{ge^{\prime}}=\mu_{ge}=\mu_{e^{\prime}f}$ etc.). The
corresponding TDHF expression is: $\left(  \Omega_{2}-\bar{\omega}%
_{fg}\right)  ^{-1} \left(  \Omega_{3}-\omega_{e^{\prime}g}\right)  ^{-2}. $
The different $\Omega_{3}$ dependencies reflect different numbers of
resonances with different FWHM along the $\Omega_{3}$ axis. The double
resonance in TDHF is split into two resonances in the exact expression.

Fig.~2C shows the signal from two coupled quantum dots, each hosting one
valence and one conduction orbital. The $S_{CI}$ signal contains a rich peak
structure, reflecting the 4 (10) many body levels in the single- (double-)
excited manifold. Again, the TDHF method misses many peaks. In this case,
unlike the two previous systems, TDHF does not show all possible resonances
along the $\Omega_{2}$ axis. This is because one of the single-excited states
($e_{1}$) is not optically allowed. In TDHF any $f$ state, constructed as a
direct product of $e_{1}$ with another state $e_{i}$ ($i=1,\ldots,4$) is
forbidden, thus we have only $6$ resonances. In the exact calculation, the $f$
states are not direct products, so we see all possible resonances along
$\Omega_{2}$. The differences between the TDHF and exact spectra in all these
examples, illustrate the sensitivity of the proposed signal to the correlated wavefunction.

Computing electron correlation effects, which are neglected by HF theory,
constitutes a formidable challenge of many-body theory. Each higher-level
theory for electron correlations \cite{Giuliani,Marques} is expected to
predict a distinct two-dimensional signal, which will reflect the accuracy of
its energies and many-body wavefunctions. The proposed technique thus offers a
direct experimental test for the accuracy of the energies as well as the
many-body wavefunctions calculated by different approaches. Time dependent
density functional theory TDDFT within the adiabatic approximation extends
TDHF to better include exchange and correlation effects \cite{Marques,Onida}%
.\ However, the two are formally equivalent and yield a similar excited-state
structure \cite{Berman}. The two-dimensional peak pattern of TDDFT will suffer
from the same limitations of TDHF.

We can summarize our findings as follows: at the HF level which assumes
\textit{independent electrons}, the $S_{CI}$ signal vanishes due to
interference. TDHF (or TDDFT) goes one step further and provides a picture of
\textit{independent} \textit{transitions} (quasiparticles). Here the signal no
longer vanishes, but shows a limited number of peaks. When correlation effects
are fully incorporated, the many-electron wavefunctions become superpositions
of states with different numbers and types of e-h pairs. The $\Omega_{2}$ and
$\Omega_{3}$ axes will then contain many more peaks corresponding to all many
body states (in the frequency range spanned by the pulse bandwidths), which
project into the doubly-excited states. Thus, along $\Omega_{2}$ the peaks
will be shifted, reflecting the level of theory used to describe electron
correlations. Along $\Omega_{3}$, the effect is even more dramatic and new
peaks will show up corresponding to splittings between various levels. This
highly-resolved two dimensional spectrum provides a invaluable direct
dynamical probe of electron correlations (both energies and wavefunctions).

Signals obtained from a similar pulse sequence, calculated for electronic
transitions in molecular aggregates\cite{mukamelannrev2000} and molecular
vibrations\cite{ZhuangMukamel2005} show the role of coupling between Frenkel
excitons. A conceptually related NMR technique known as double quantum
coherence reveals correlation effects among spins. The technique showed
unusual sensitivity for weak couplings between spatially remote spins and has
been used to develop new MRI imaging techniques \cite{Richter,MukamelCPL}.
Here we have extended this idea to all many-electron systems. The proposed
technique should apply to molecules, atoms, quantum dots and highly correlated
systems such as superconductors. It has been recently demonstrated that
two-exciton couplings can be controlled in onion-like semiconductor
nanoparticles with a core and an outer shell made of different materials
\cite{Klimov}. \ Nonlinear spectroscopy of the kind proposed here could
provide invaluable insights into the nature of such two-exciton states.

\begin{acknowledgments}
This research was supported by the National Science Foundation Grant No.
CHE-0446555 and the National Institutes of Health Grant No. GM59230.
\end{acknowledgments}

\newpage

\textbf{Figure Captions}

\vspace{1cm}

Fig.~1 Left: many-body states connected by transitions dipoles, including the
ground state $g$, the manifold of single e-h pairs $e$ and the manifold of
two-pair states $f$. Right: the two Feynman diagrams contributing to the
correlation-induced signal $S_{CI}$ (Eq~\ref{Eq.SOS_resp}). $t_{i}$ are the
time delays between laser pulses. For independent electrons $\omega
_{fe}=\omega_{e\prime g}$ and diagrams of type (i) and (ii) cancel in pairs.

Fig.~2 Absolute value of the exact and the TDHF $S_{CI}$ signals for three
model systems. Energies on the axes are referenced to the carrier frequency,
which excites interband transitions. Each system has $N(N+1)/2$ doubly-excited
levels, where $N$ is the number of single-excited levels. (A) 2 orbital system
with $N=1$ ($V^{eh}=194.4$cm$^{-1}$), (B) 3 orbital system with $N=2$
($V^{eh}=194.4$cm$^{-1}$), (C) 2 coupled systems of type (A), but with
different gaps, $V^{eh}=193.9$ cm$^{-1}$ for one dot and $78.1$cm$^{-1}$ for
the other and $t=59.2$cm$^{-1}$, $N=4$. The corresponding orbitals and
splittings are given schematically below each panel. For systems (A) and (B),
the TDHF misses half of the resonances along $\Omega_{3}$, while for (C) it
misses 4 out of 10 resonances along $\Omega_{2}$.

\end{document}